\documentclass[twocolumn,showpacs,preprintnumbers,superscriptaddress,amsmath,amssymb]{revtex4}
\newcommand{\pT}{\ensuremath{p_T}}
\newcommand{\snn}{\ensuremath{\sqrt{s_{NN}}}}
\newcommand{\raa}{\ensuremath{R_{AA}}}
\newcommand{\npart}{\ensuremath{ N_{\rm part}}}

\usepackage{graphicx}
\usepackage{dcolumn}
\usepackage{bm}
\usepackage{color}

\begin{document}

\preprint{Version 7.0}

\title{Identified high-\pT~spectra in Cu+Cu collisions at \snn=200~GeV}

\date{\today}

\affiliation{Argonne National Laboratory, Argonne, Illinois 60439, USA}
\affiliation{University of Birmingham, Birmingham, United Kingdom}
\affiliation{Brookhaven National Laboratory, Upton, New York 11973, USA}
\affiliation{University of California, Berkeley, California 94720, USA}
\affiliation{University of California, Davis, California 95616, USA}
\affiliation{University of California, Los Angeles, California 90095, USA}
\affiliation{Universidade Estadual de Campinas, Sao Paulo, Brazil}
\affiliation{University of Illinois at Chicago, Chicago, Illinois 60607, USA}
\affiliation{Creighton University, Omaha, Nebraska 68178, USA}
\affiliation{Czech Technical University in Prague, FNSPE, Prague, 115 19, Czech Republic}
\affiliation{Nuclear Physics Institute AS CR, 250 68 \v{R}e\v{z}/Prague, Czech Republic}
\affiliation{University of Frankfurt, Frankfurt, Germany}
\affiliation{Institute of Physics, Bhubaneswar 751005, India}
\affiliation{Indian Institute of Technology, Mumbai, India}
\affiliation{Indiana University, Bloomington, Indiana 47408, USA}
\affiliation{University of Jammu, Jammu 180001, India}
\affiliation{Joint Institute for Nuclear Research, Dubna, 141 980, Russia}
\affiliation{Kent State University, Kent, Ohio 44242, USA}
\affiliation{University of Kentucky, Lexington, Kentucky, 40506-0055, USA}
\affiliation{Institute of Modern Physics, Lanzhou, China}
\affiliation{Lawrence Berkeley National Laboratory, Berkeley, California 94720, USA}
\affiliation{Massachusetts Institute of Technology, Cambridge, MA 02139-4307, USA}
\affiliation{Max-Planck-Institut f\"ur Physik, Munich, Germany}
\affiliation{Michigan State University, East Lansing, Michigan 48824, USA}
\affiliation{Moscow Engineering Physics Institute, Moscow Russia}
\affiliation{City College of New York, New York City, New York 10031, USA}
\affiliation{NIKHEF and Utrecht University, Amsterdam, The Netherlands}
\affiliation{Ohio State University, Columbus, Ohio 43210, USA}
\affiliation{Old Dominion University, Norfolk, VA, 23529, USA}
\affiliation{Panjab University, Chandigarh 160014, India}
\affiliation{Pennsylvania State University, University Park, Pennsylvania 16802, USA}
\affiliation{Institute of High Energy Physics, Protvino, Russia}
\affiliation{Purdue University, West Lafayette, Indiana 47907, USA}
\affiliation{Pusan National University, Pusan, Republic of Korea}
\affiliation{University of Rajasthan, Jaipur 302004, India}
\affiliation{Rice University, Houston, Texas 77251, USA}
\affiliation{Universidade de Sao Paulo, Sao Paulo, Brazil}
\affiliation{University of Science \& Technology of China, Hefei 230026, China}
\affiliation{Shandong University, Jinan, Shandong 250100, China}
\affiliation{Shanghai Institute of Applied Physics, Shanghai 201800, China}
\affiliation{SUBATECH, Nantes, France}
\affiliation{Texas A\&M University, College Station, Texas 77843, USA}
\affiliation{University of Texas, Austin, Texas 78712, USA}
\affiliation{Tsinghua University, Beijing 100084, China}
\affiliation{United States Naval Academy, Annapolis, MD 21402, USA}
\affiliation{Valparaiso University, Valparaiso, Indiana 46383, USA}
\affiliation{Variable Energy Cyclotron Centre, Kolkata 700064, India}
\affiliation{Warsaw University of Technology, Warsaw, Poland}
\affiliation{University of Washington, Seattle, Washington 98195, USA}
\affiliation{Wayne State University, Detroit, Michigan 48201, USA}
\affiliation{Institute of Particle Physics, CCNU (HZNU), Wuhan 430079, China}
\affiliation{Yale University, New Haven, Connecticut 06520, USA}
\affiliation{University of Zagreb, Zagreb, HR-10002, Croatia}

\author{B.~I.~Abelev}\affiliation{University of Illinois at Chicago, Chicago, Illinois 60607, USA}
\author{M.~M.~Aggarwal}\affiliation{Panjab University, Chandigarh 160014, India}
\author{Z.~Ahammed}\affiliation{Variable Energy Cyclotron Centre, Kolkata 700064, India}
\author{A.~V.~Alakhverdyants}\affiliation{Joint Institute for Nuclear Research, Dubna, 141 980, Russia}
\author{B.~D.~Anderson}\affiliation{Kent State University, Kent, Ohio 44242, USA}
\author{D.~Arkhipkin}\affiliation{Brookhaven National Laboratory, Upton, New York 11973, USA}
\author{G.~S.~Averichev}\affiliation{Joint Institute for Nuclear Research, Dubna, 141 980, Russia}
\author{J.~Balewski}\affiliation{Massachusetts Institute of Technology, Cambridge, MA 02139-4307, USA}
\author{L.~S.~Barnby}\affiliation{University of Birmingham, Birmingham, United Kingdom}
\author{S.~Baumgart}\affiliation{Yale University, New Haven, Connecticut 06520, USA}
% ...

\collaboration{STAR Collaboration}\noaffiliation

\begin{abstract}
We report new results on identified (anti)proton and charged pion spectra
at large transverse momenta (3$<$\pT$<$10~GeV/$c$) from Cu+Cu collisions at \snn~=~200~GeV
using the STAR detector at the Relativistic Heavy Ion Collider (RHIC).
This study  explores the system size dependence of  two novel features
observed at RHIC with  heavy ions: the hadron suppression at high-\pT~and
the anomalous baryon to meson enhancement at intermediate transverse momenta.
Both phenomena could be attributed to the creation of a new form of QCD matter.
The results presented here bridge the system size gap between
the available $pp$ and Au+Au data, and allow the detailed  exploration
for the on-set of the novel features.  Comparative analysis of all
available 200~GeV data indicates that the system size is a major factor
determining both the magnitude of the hadron spectra suppression at large
transverse momenta and the relative baryon to meson enhancement.
\end{abstract}


\maketitle

Differential studies of identified particle production in nucleus-nucleus collisions
provide an experimental means to probe the different stages of the collision evolution and explore the properties of the created medium.  Spectral measurements at high transverse momenta are of special interest, for the following reasons. In elementary collisions, hard partonic scatterings are known to produce jets of particles originating from the fragmentation of a high-\pT~quark or gluon.  The spectral distributions of
particles in transverse momentum from such interactions are measured experimentally and are reasonably well understood in terms of Next-to-Leading Order (NLO) pQCD calculations~\cite{cite:STAR_Id200dAupp}.
These hard scatterings occur in heavy-ion collisions as well,
but their resulting distributions are found to be modified due
to interactions with the medium and the resulting energy loss.  Thus, understanding modifications to
the high-\pT~particle distributions is an important step towards understanding the partonic energy loss mechanisms within the
medium~\cite{cite:Theory_Vitev}.

To study the effects of parton-medium interaction on particle production in heavy-ion collisions we compare the production cross-sections measured in AA to the equivalent measurements in $pp$ collisions. Following the expectation that the particle production in heavy-ion collisions at high-\pT~is determined by the number of binary nucleon-nucleon inelastic collisions we define the nuclear modification factor, $R_{AA}$, as the ratio (Eq.~(\ref{eqn:RAA})) of particle yields measured in AA to the cross-sections measured in $pp$ collisions scaled by the corresponding number of independent nucleon-nucleon collisions $N_{\rm bin}^{AA}$. We obtain $N_{\rm bin}^{AA}$ from a Monte Carlo Glauber model calculation~\cite{cite:STAR_SpectralSuppression}. For the unmodified particle production in AA collisions $R_{AA}$ is exactly unity, whilst $R_{AA}$$<$1 indicates {\it suppression} and $R_{AA} > 1$ {\it enhancement}.

\begin{equation}
 \raa = \frac{\sigma_{NN}^{\rm inel}}{N_{\rm bin}^{AA}} \frac{d^{2}N_{AA}/dyd\pT}{d^{2}\sigma_{pp}/dyd\pT} \,.
\label{eqn:RAA}
\end{equation}
\vspace{5pt}

\noindent
The $R_{AA}$ measured in $d$+Au and peripheral Au+Au collisions exhibits an enhanced particle production which is believed to occur due to multiple nucleon scatterings within the colliding nuclei. This ``initial'' state effect is known as the Cronin effect~\cite{cite:STAR_Id200dAupp,cite:STAR_ToF,cite:CroninEffect}.
Meanwhile, in central Au+Au collisions, $R_{AA}$ at high-$\pT$ indicates that the particle production is strongly suppressed (by about a factor of 5)~\cite{cite:STAR_SpectralSuppression,cite:STAR_BackToBack}. This ``final'' state effect has been attributed to the partonic energy loss in an opaque colored medium~\cite{cite:STAR_dAuBackToBack}. However, neither of the two effects is sufficiently understood and both require further experimental and theoretical study.
The differential analysis presented here  explores the system size effects on  parton propagation through the medium to  further evaluate  the mechanisms of  parton/medium
interactions.    
\begin{figure*}[t]
\includegraphics[width=0.95\textwidth]{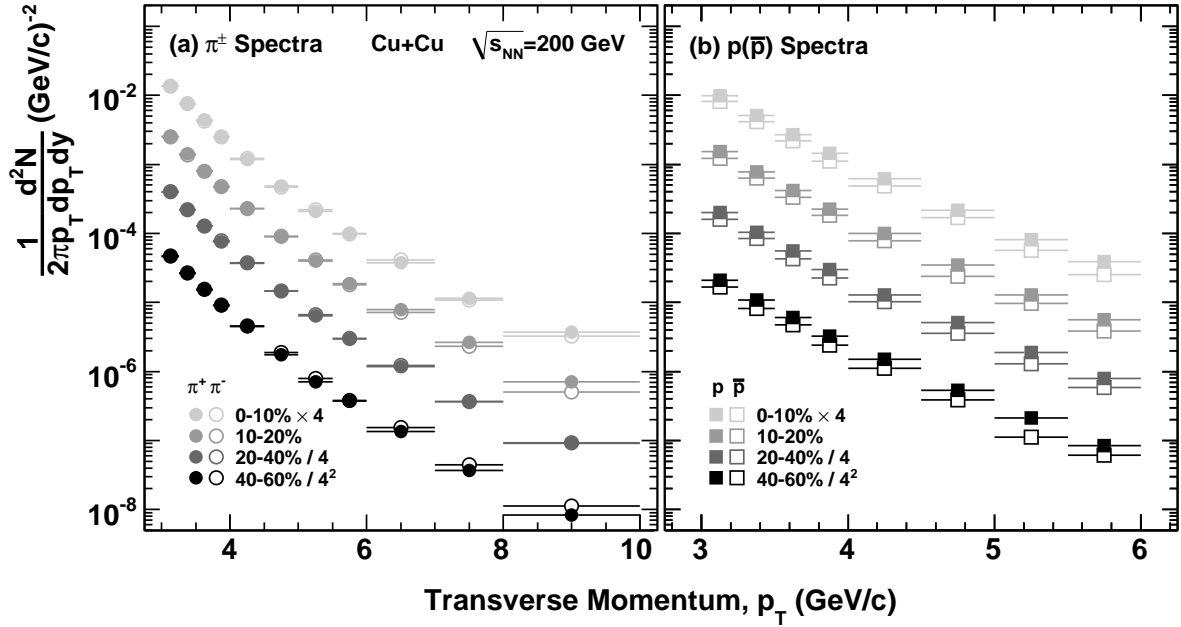}
\caption{\label{fig:Spectra} Transverse momentum spectra of
pions $(a)$ and protons $(b)$ produced in Cu+Cu collisions
at \snn=200~GeV.  Data are presented for four centrality classes:
0-10\%, 10-20\%, 20-40\% and 40-60\%. Closed and open symbols are used for particles and antiparticles, respectively.  For clarity, data are
separated by powers of four.}
\end{figure*}

To provide additional constraints and systematic understanding of the measurements in very light ($d$+Au) and heavy (Au+Au) collision systems we present the key studies at the intermediate (Cu+Cu) system at the same incident energy ($\sqrt{s_{NN}} = 200$~GeV), bridging the gap between the two extremes.
These measurements may provide quantitative understanding of the partonic energy loss and its system size dependence.
In addition, it is expected that the identified particle measurements provide information on color-charge effects within the mechanism of jet quenching.
Although experimental discrimination between quark and gluon jet fragmentation on event-by-event basis is difficult, it can be addressed statistically by exclusive analysis of proton (or baryon) and pion (meson) production.
We are utilizing the idea that 
the baryon to meson ratio is found higher in gluon jets compared to quark jets~\cite{4Macl}.
Identified proton and pion measurements from $pp$ collisions
concur with this picture~\cite{cite:STAR_Id200dAupp,cite:STAR_Id200AuAu},
as well as direct measurements of baryon and meson production in
quark and gluon jets~\cite{cite:DELPHI_Id_in_QGJets}.
Thus, identified particle measurements at high-\pT~can then be used to analyze gluon and quark propagation through the medium and
to probe the color-charge differences of energy loss~\cite{cite:STAR_Id200AuAu,cite:Theory_Vitev,cite:STAR_62Id,cite:STAR_BedangaProc}.

Additionally, systematic studies of identified particle production in Cu+Cu can shed new light on the anomalous enhancement of baryons with respect to mesons observed at intermediate transverse momenta ($2$$<$\pT$<$6~GeV/$c$) in Au+Au collisions.
This enhancement is not consistent with the extrapolated values from the measurements in $pp$ collisions and cannot be explained by cold nuclear matter effects.
At present, the preferred baryon over meson production at intermediate \pT~could be described by two very different considerations. The first model assumes  coalescence and recombination, which demands a shift of baryon yields to higher momenta relative to meson yields~\cite{cite:HwaRecombination}.
The second model evokes an interplay of the flow effects in the radially expanding medium with the jet fragmentation~\cite{SplusH}.

In this paper identified (anti)proton and charged pion spectra  are
systematically explored with regard to the system size with the smaller
Cu+Cu colliding system  at \snn=~200~GeV. The centrality dependence of
high-\pT~hadron production and the \pT~dependence of baryon to meson
ratios in Cu+Cu data are compared to the Au+Au system as well as to $pp$
collisions at the same energy.
This allows  gaining a greater understanding of {\it peripheral}
collisions.  The size of Cu nuclei is ideally suited to
explore the turn-on of the high-\pT~suppression bridging the gap
between $pp$, $d$+Au and peripheral Au+Au data in terms of system size
and nuclear matter.

The Cu+Cu data used in this analysis were recorded by the STAR
experiment during Run 5  at RHIC.  Here, the minimum bias trigger
was based on the combined signals from the Beam-Beam Counters at
forward rapidity ($3.3$$<$$|\eta|$$<$$5.0$) and the Zero-Degree Calorimeters, located at $\pm 18$~m from the nominal interaction point~\cite{refTrigZDC}.  
In total, 23~million events comprise this data set.  
Based on the charged track multiplicity recorded in the Time Projection Chamber (TPC) and Glauber MC model calculations,  the data are divided into four centrality bins corresponding to 0-10\%, 10-20\%, 20-40\% and 40-60\% of fractional cross-section ($\sigma/\sigma_{\rm geom}$) bins.
In order to remove as many background tracks as possible, tracks
which intercept the measured collision vertex within 1~cm (distance of closest
approach) were retained with a minimum of 25 (out of 45)
TPC trajectory points forming each track.

Within the STAR experiment, particle identification at low-\pT~is
attained by use of the ionization energy loss ($dE/dx$) in the
TPC~\cite{cite:STAR_TPC_NIM}.
For low momentum particles, below $1$~GeV$/c$, a clear
mass separation is observed allowing the identification of
$\pi^{\pm}$, K$^{\pm}$ and (anti)protons.  In the
intermediate-\pT~region ($1$$<$\pT$<$$3$~GeV$/c$) the TPC is no
longer directly usable by itself, as all particles, independent
of mass, are minimum ionizing.  
For the purpose of this paper we identify pions, kaons, protons and anti-protons at higher momenta (\pT$>3$~GeV$/c$) on a statistical basis utilizing the relativistic rise of the ionization energy loss in the TPC.
For a given slice in transverse
momentum, a distinctly non-single-Gaussian shape is observed,
discussed in detail in~\cite{cite:STAR_dEdx_NIM}, representing
the normalized deviations from different energy loss trends of $\pi$, K and protons. The quantity
used to express the energy loss is a normalized distribution,
$n_{\sigma}$ defined in Eq.~(\ref{eqn:Nsigma}), which accounts for
the theoretical expectation ($B_{\pi}$, known as a Bichsel parameterization) and the resolution of the
TPC for pions ($\sigma_{\pi}$).

\begin{equation}
n_{\sigma}={\rm log}((dE/dx)/(B_{\pi}))/\sigma_{\pi}
\label{eqn:Nsigma}
\end{equation}

\noindent The resultant distribution in each transverse momentum range
is fit with a six-Gaussian function (one per particle-species/charge).
The Gaussian widths are considered to be the same, independent of
particle type, and single-Gaussian centroids are defined by the
theoretical expectations constrained by the identified proton and pion measurements
from topologically reconstructed weakly decaying particle yields~\cite{cite:STAR_dEdx_NIM}.
Further details of the particle identification technique can be found in
Refs.~\cite{cite:STAR_Id200dAupp,cite:STAR_ToF}.

Raw data yields are corrected for single-track inefficiencies evaluated via
Monte-Carlo tracks embedded into real data events.  
We define single-track efficiency as the fraction of Monte Carlo tracks embedded into real Cu+Cu events that have been reconstructed.
The efficiencies are derived for each different event multiplicity bin and particle species.
For high-\pT~tracks (\pT$>2$~GeV$/c$) in 200~GeV Cu+Cu events, the efficiency
is found to be ~85\% on average, with a weak ($<$10\%) centrality and \pT~dependence.
In the analysis, pion and (anti)proton abundances  are extracted from the $n_{\sigma}$ distribution
using the finely calibrated centroid positions.  The derived kaon yields
are then smoothed to reduce statistical fluctuations, using a Levy~\cite{cite:STAR_Id200dAupp} fit.
These assumed kaon spectra are then used for a final fit to determine
the pion and proton yields.  The systematic uncertainties from this
procedure, on the spectra, are 2-10\% for pions and 5-11\% for protons, decreasing smoothly with \pT~in the measured range.
An analysis solving simultaneous equations to assumed
pion, kaon and (anti)proton distributions (bin counting), derived
results that are 5-10\% (5-20\%) higher for pions (protons).  This
difference is the dominant systematic uncertainty on particle spectra.  An additional
systematic error of 5\% resultant from the uncertainty in the single-particle
efficiency determination is added in quadrature.  The total systematic error for pion spectra
ranges from 9\% (at 3~GeV/$c$) to 13\% (at 10~GeV/$c$).  For protons,
the error ranges from 21\% (at 3~GeV/$c$) to 23\% (at~6~GeV/$c$).  These uncertainties are similar to the earlier evaluation from Au+Au data analysis~\cite{cite:STAR_Id200AuAu}.
Systematic uncertainties from other possible sources such as the momentum resolution (studied by embedding) and the uncertainty in determination of the centroid position (within the particle identification procedure) are negligible.
The corrected transverse momentum spectra for $\pi^{\pm}$ and
(anti)protons at \snn=200~GeV are shown in
Fig.~\ref{fig:Spectra} $(a)$ and $(b)$ respectively.  The reach in transverse momentum is limited
only by the available statistics.  Additional \pT~reach
for pion identification is due to a larger separation from the kaon peak
($\Delta\sigma_{\pi K}\sim 2\sigma$, $\Delta\sigma _{pK}\sim 1\sigma$).

\begin{figure*}[t]
\includegraphics*[height=8.5cm]{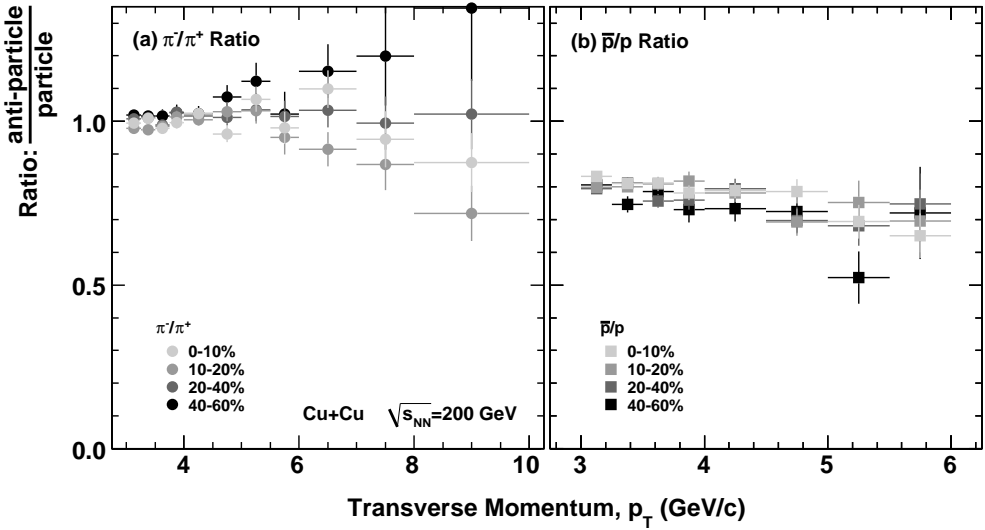}
\caption{\label{fig:Ratios} Anti-particle to particle ratios, as a function
of transverse momentum for pions $(a)$ and protons $(b)$.
Data for the four centrality classes show little centrality dependence.
Errors are statistical only.}
\end{figure*}

Figure~\ref{fig:Ratios} shows 
$\pi^-/\pi^+$ and $\bar{p}/p$ ratios
in Cu+Cu data at \snn=200~GeV for the four centrality bins.  The data
show no systematic trends versus centrality within uncertainties, and a weak (if any) decreasing
$\overline{p}/{p}$ with transverse momentum (as observed in
Au+Au collisions~\cite{cite:STAR_Id200AuAu}).  Thus, to improve the
statistical uncertainties in the following discussion, data are averaged
over particle charge.

\begin{figure*}[t]
\includegraphics*[height=7.5cm]{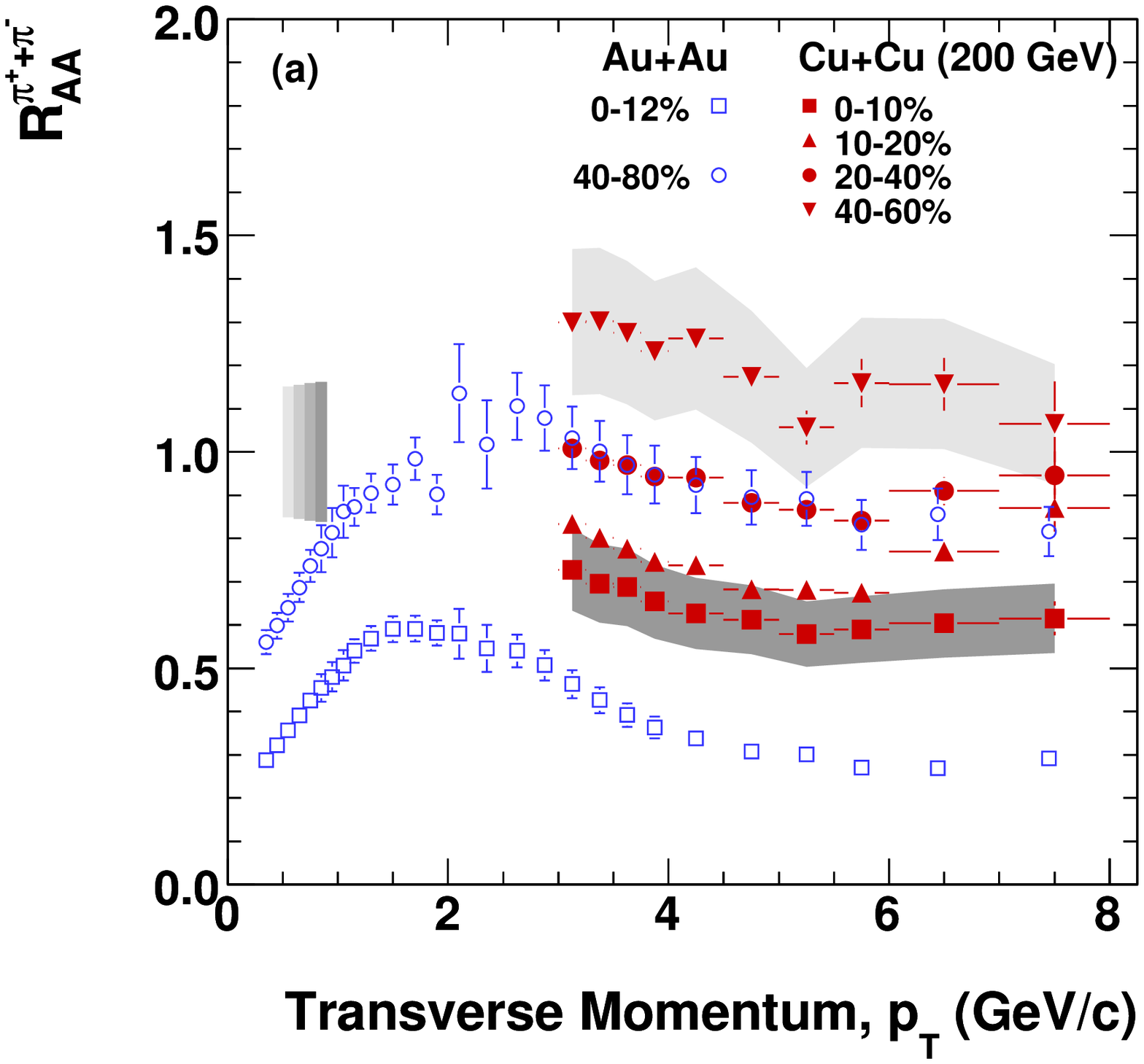}
\includegraphics*[height=7.5cm]{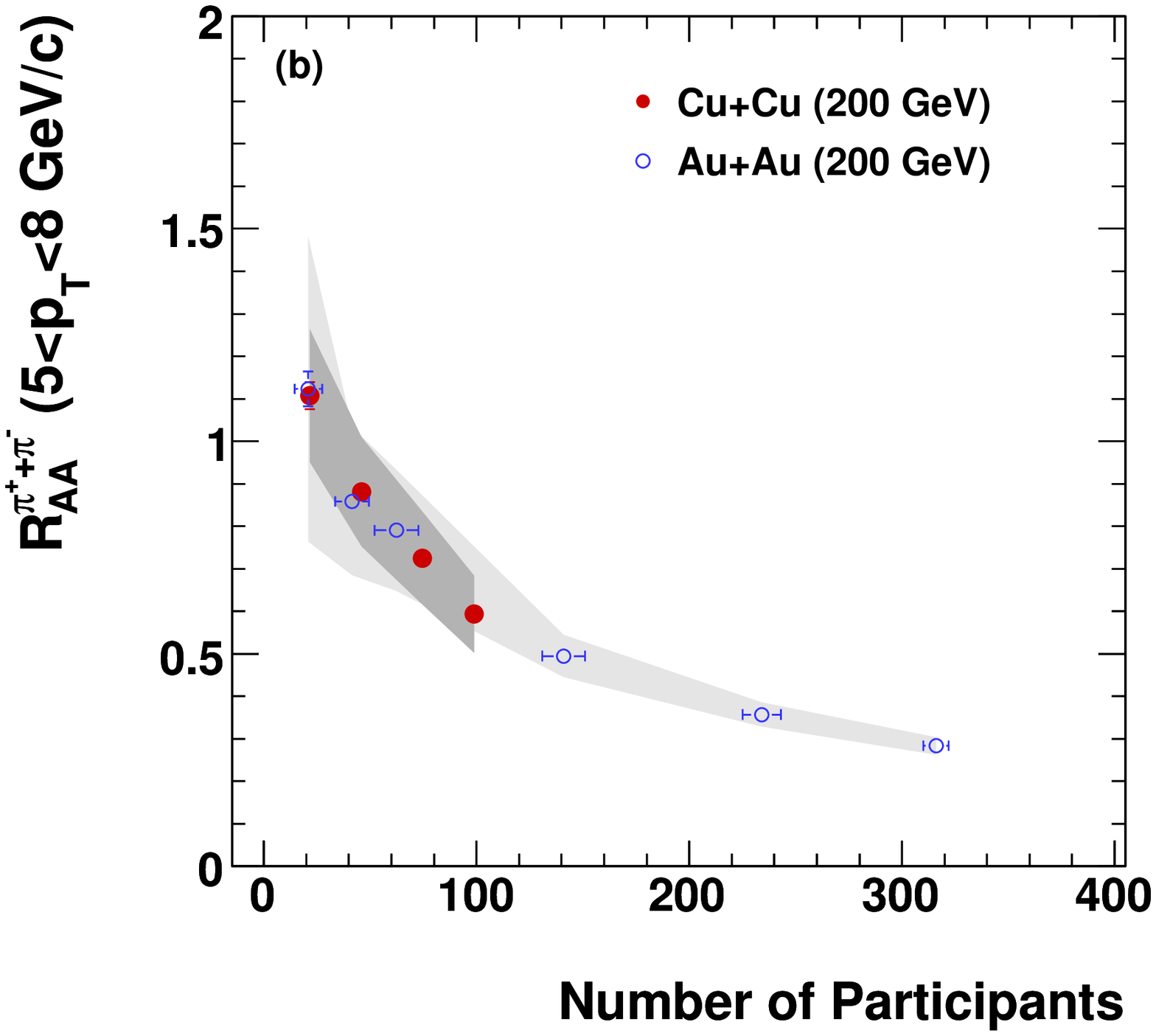}
\caption{\label{fig:PionRAA} (Color online) {\it (a)} Nuclear modification factor, \raa, for
charged pions ($\pi^++\pi^-$) in Cu+Cu (filled symbols) and Au+Au (open symbols)
collisions at \snn=200~GeV.
Error bands are shown for most peripheral and most central Cu+Cu data to represent evolution of the systematic uncertainties for this dataset. Error boxes at \raa=1 represent Cu+Cu
scale uncertainties due to the number of collisions and from $pp$ spectra normalization.  {\it (b)} Integrated pion \raa~over the range 5$<$\pT$<$8~GeV/{\em c} versus \npart.
The bands represent the systematic uncertainty on ratios.  An additional
scale error due to $pp$ normalization ($\sim$14\%) is not shown.
}
\end{figure*}

The spectral data alone can convey only a limited message.  To delve into properties of the resultant data, ratios are taken.  The first such ratio is termed the nuclear modification factor (\raa),
defined in Eq.~(\ref{eqn:RAA}).  
We find that the pion spectra are suppressed
in the most central ({\em head-on}) Cu+Cu data at \snn=200~GeV (Fig.~\ref{fig:PionRAA}).  For the peripheral ({\em glancing})
collisions, a small enhancement is observed.  
To expose the features of the modifications of the hadron spectra in Cu+Cu and Au+Au collisions we study \raa~as a function of the amount of matter participating in the collisions. For both systems \raa~is evaluated within several fractional cross-section bins and as a function of the number of participating nucleons.
Figure ~\ref{fig:PionRAA}~({\it a}) shows the results of this comparative analysis
using the most central 0-12\% (open squares) and mid-peripheral 40-60\% (open circles) Au+Au data.  
For the most central events the suppression level
is found to be different between the systems.  The resultant
spectra from Au+Au collisions are more suppressed than in Cu+Cu data.
According to the Glauber calculation the
mid-central (20-40\%) Cu+Cu collisions (closed circles) and
mid-peripheral (40-80\%) Au+Au data (open circles)  have similar numbers of participating nucleons (see Appendix A for details).   
For this selection of centralities within the two systems we find that numerical values of \raa~agree within the uncertainties. This agreement suggests a correlation of the suppression with the initial volume of the collision system.

\begin{figure*}[t]
\includegraphics*[height=7.5cm]{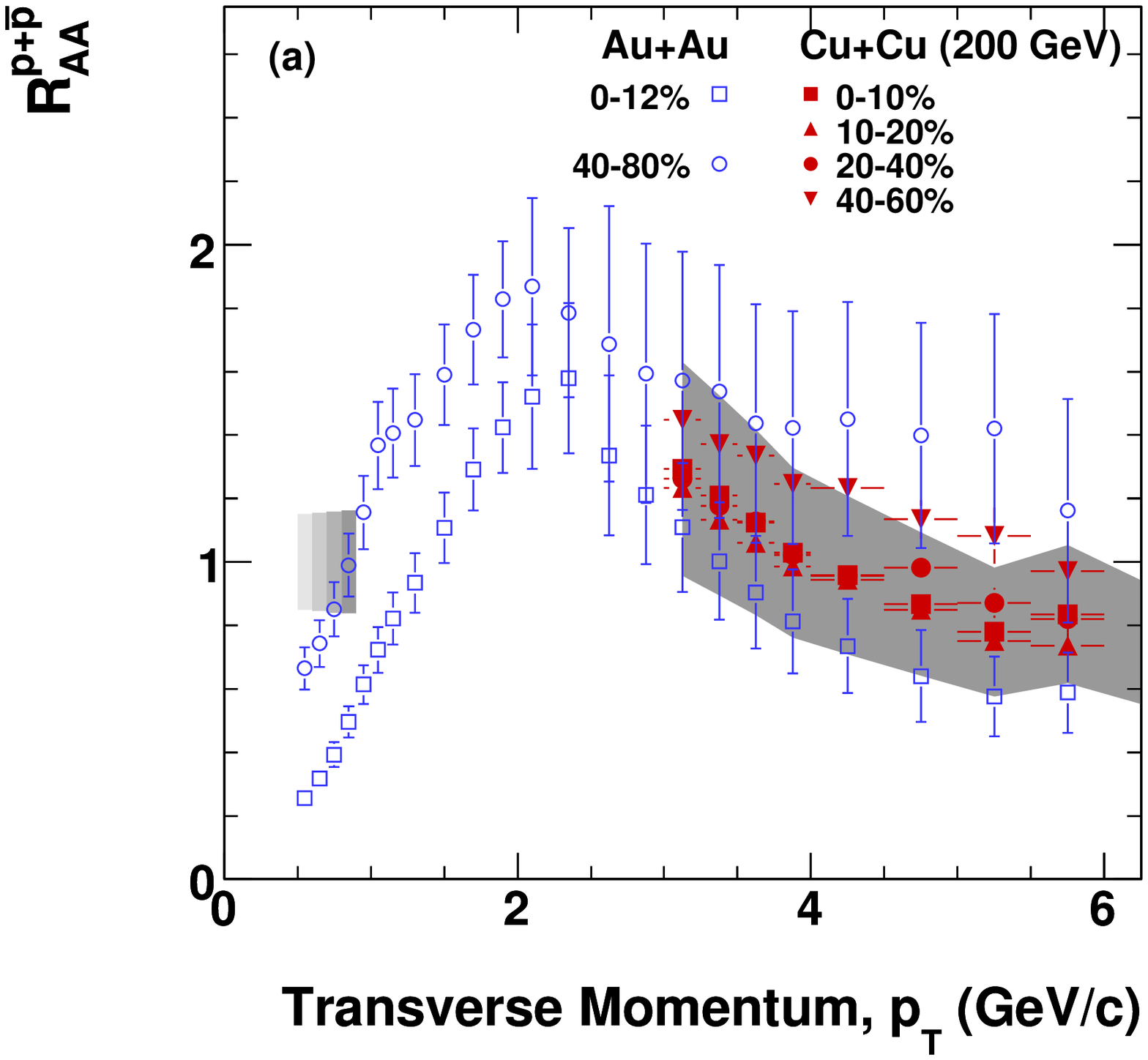}
\includegraphics*[height=7.5cm]{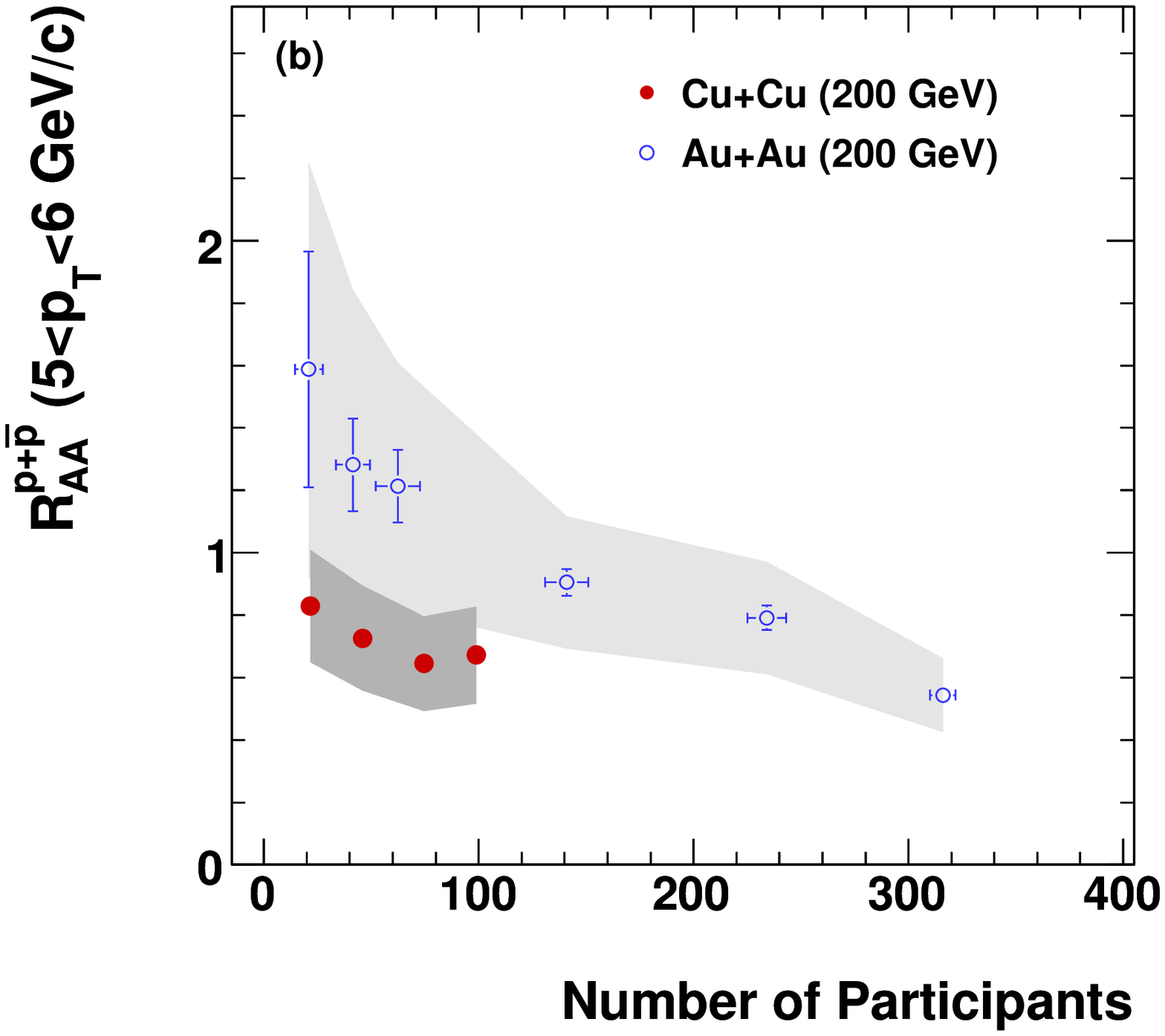}
\caption{\label{fig:ProtonRAA} (Color online) {\it (a)} Nuclear modification factor, \raa, for
protons and anti-protons ($p+\bar{p}$) in Cu+Cu (filled symbols) and Au+Au (open symbols)
collisions at \snn=200~GeV.
Error band is shown for most central Cu+Cu data to represent characteristic systematic uncertainties for Cu+Cu data. Error boxes at \raa=1 represent Cu+Cu scale uncertainties due to the number of collisions and from $pp$ spectra normalization.
{\it (b)} Integrated (anti)proton \raa~over the range 5$<$\pT$<$6~GeV/{\em c} versus \npart.
The bands represent the systematic uncertainty on ratios.  An additional
scale error due to $pp$ normalization (about 14\%) is not shown.
}
\end{figure*}

In Fig.~\ref{fig:PionRAA}~({\it b}) we present the \pT~averaged \raa~for pions (5$<$\pT$<$8~GeV/$c$) as a function of the number of participating nucleons calculated for both Cu+Cu and Au+Au collisions.
The agreement between Au+Au (open circles)
and Cu+Cu (closed circles) is striking and demonstrates that the nuclear modification factor for pions is a smooth function of the number of participating nucleons (independent of the collision system).

Similarly, we explore the systematics of baryon production in Cu+Cu and Au+Au systems by comparing the \raa~for protons and anti-protons. Figure~\ref{fig:ProtonRAA}~({\it a}) shows
the \raa~distributions averaged over $p$ and $\bar{p}$ for four centrality bins of Cu+Cu events.
The data at hand  does not differentiate if collision volume ($N_{\rm part}$) or fractional cross-section effects are driving the high-\pT~suppression for baryons due to the larger systematic uncertainties for (anti)proton measurements.
Nevertheless, we observe that  proton production in the peripheral Cu+Cu events is consistent with binary scaling expectations, and the suppression is setting in as one progresses from the peripheral to central events.
An overall  similar centrality dependence was observed between the Au+Au and Cu+Cu data at the same energy (see Fig.~\ref{fig:ProtonRAA}~({\it b})), albeit Cu+Cu integrated \raa~values seem lower than the respective Au+Au data points. We emphasize that the systematic errors are uncorrelated between the systems, and both measurements are similar within the experimental uncertainties.

The similarity between the different systems at the same number of participants is also evident in other aspects of
the data at lower \pT~\cite{cite:STAR_CuCu_LoPt}.  The smooth dependence of the nuclear
modification factor could be interpreted as a consequence of
medium induced energy loss of partons traversing the hot and dense
medium.  For the smaller systems sizes, either peripheral Au+Au
or Cu+Cu data, the path length traversed is smaller (on average)
than for the larger system (central Au+Au).  As observed in the
data, a smaller energy loss is thus predicted~\cite{cite:Theory_Vitev}.

\begin{figure*}[t]
\includegraphics*[height=7.5cm]{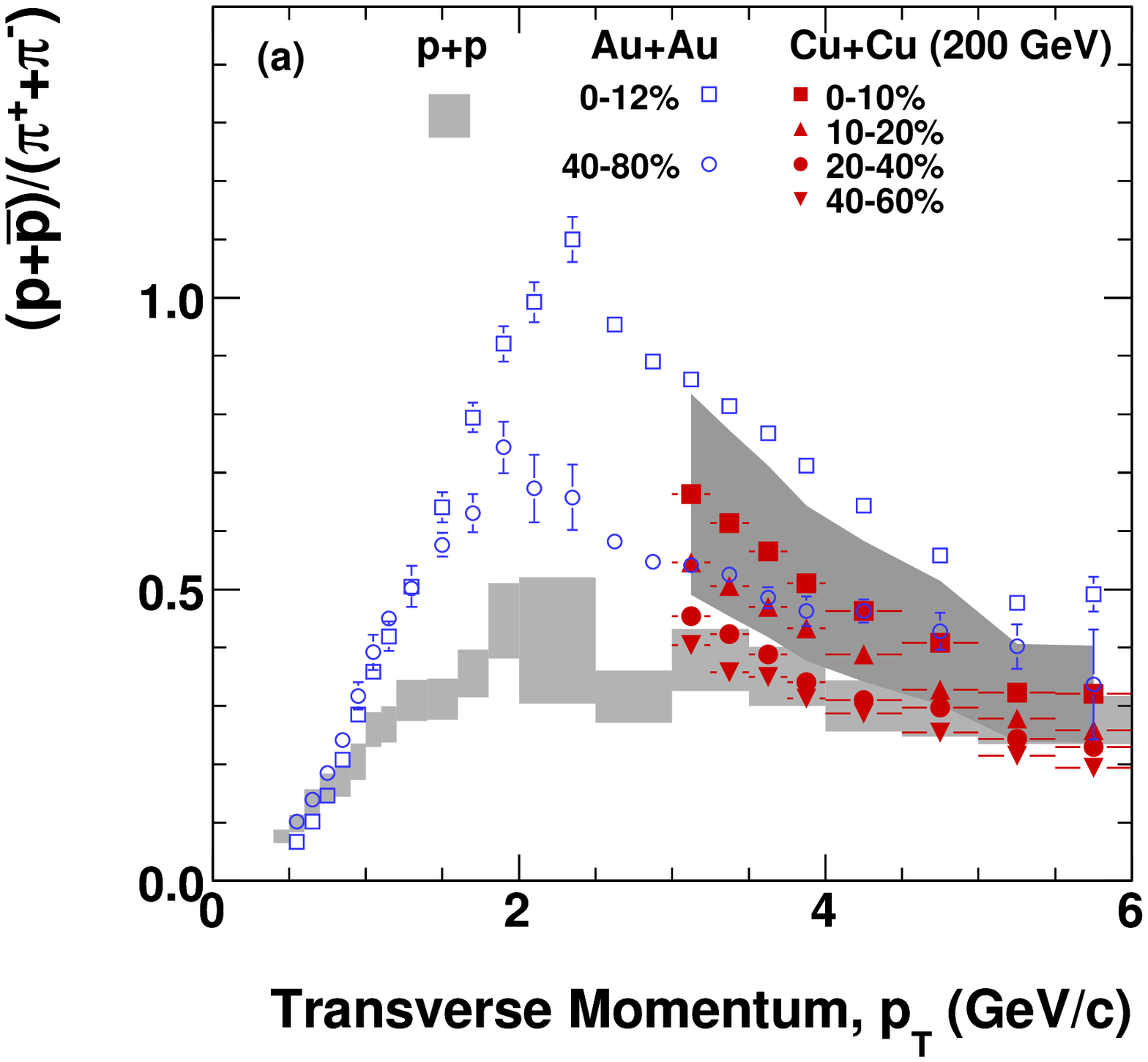}
\includegraphics*[height=7.5cm]{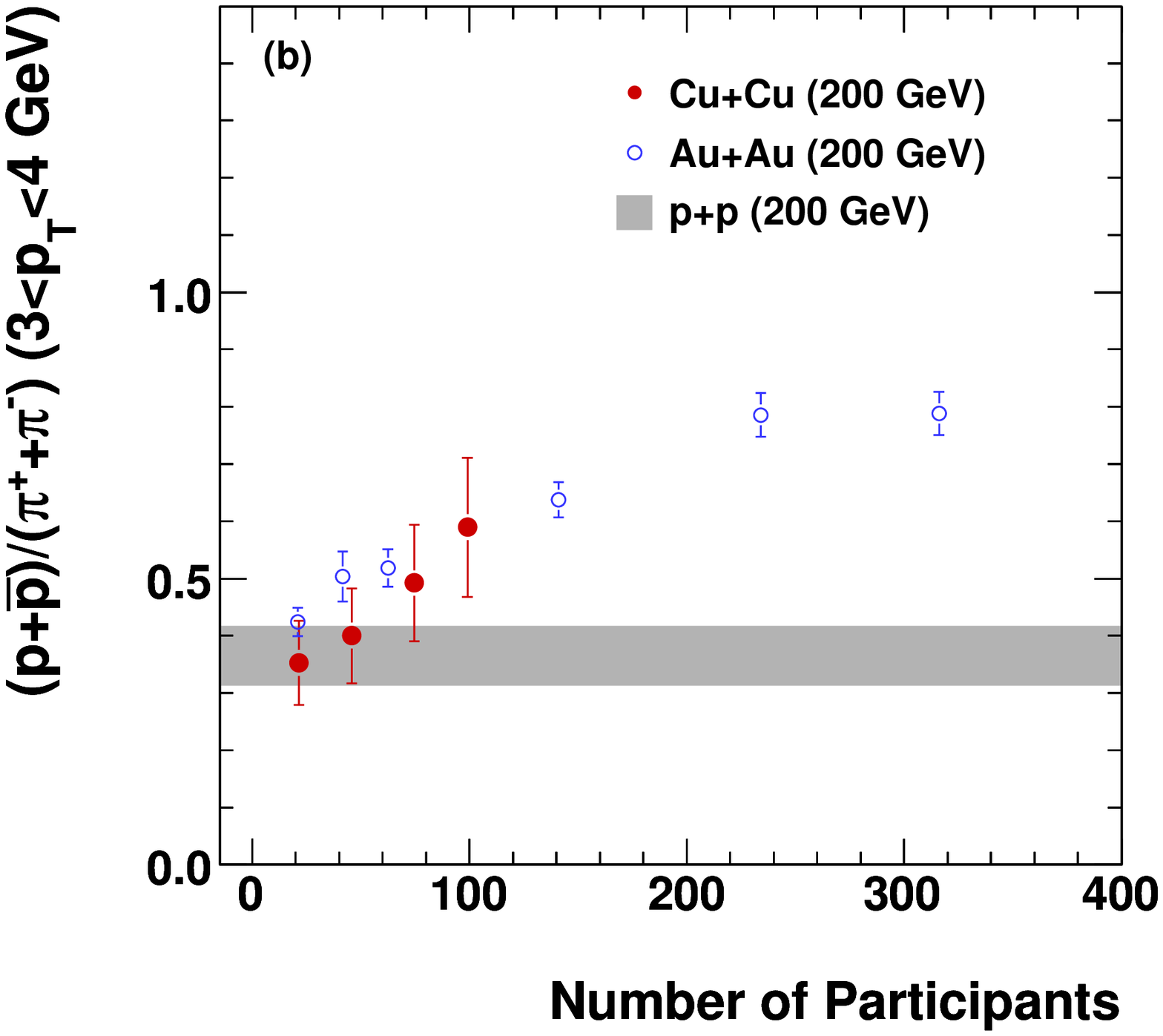}
\caption{\label{fig:BMRatio} (Color online) {\it (a)} Ratio of protons and anti-protons to
charged pions, versus transverse momentum, for Cu+Cu collisions at
\snn=200~GeV.
For clarity, only one systematic error band is shown for Cu+Cu data (most central events),
uncertainties on data in other centrality bins are similar in magnitude. Systematic errors for Au+Au data are not shown, for details see~\cite{cite:STAR_Id200AuAu}.
{\it (b)} Average $(p+\bar{p})/(\pi^{+}+\pi^{-})$ ratio for 3$<$\pT$<$4~GeV/$c$ is shown as function of centrality (\npart) for Cu+Cu and Au+Au data. Error band represents the $pp$ measurement with uncertainty.
}
\end{figure*}

Another dramatic effect observed in Au+Au data is the relative enhancement
of protons to pions in the intermediate-\pT~region as compared to $pp$ and
$e^{+}+e^{-}$ collisions~\cite{cite:STAR_Id200AuAu} as well as for other
baryon to meson ratios~\cite{cite:STAR_LambdaK0}.  This enhancement is
found to be strongly dependent on the centrality of the collision, as
illustrated in Fig.~\ref{fig:BMRatio}.
The most peripheral A+A data is shown to exhibit little or no
enhancement in this ratio, with respect to $pp$ collisions at the
same energy.  A similar increasing trend of favorable baryon
production with centrality is observed in the Cu+Cu collision
system. The peak of the enhancement is observed in the region
\pT$\sim$2~GeV/$c$ in Au+Au, at a slightly lower transverse
momentum than the range measured in this analysis. At higher
transverse momenta the enhancement over $pp$ collisions diminishes
to the level expected from vacuum fragmentation.

The baryon to meson ratio
$(p+\bar{p})/(\pi^{+}+\pi^{-})$ in Cu+Cu and Au+Au collisions shows similar trends for an equivalent number of participating nucleons.
To further quantify this observation Fig.~\ref{fig:BMRatio}~({\it b})  shows the proton to pion ratio  (for hadrons with 3$<$\pT$<$4~GeV/$c$) measured in Cu+Cu and Au+Au collisions as a function of $N_{\rm part}$.
We find that  this ratio is also sensitive to the initial volume of the collision system and exhibits the same quantitative $N_{\rm part}$ dependence irrespective of the collision system.

As discussed earlier, it is found that in the kinematic range of our measurements baryons are produced
predominantly from gluon fragmentation~\cite{AKK}.  It is thus expected that an increase in the baryon
to meson ratio in the intermediate- to high-\pT~range would be related
to gluon  sources.  To explain the presented data one could consider,
for example, that a gluon jet could be more easily propagated through
the medium than a quark jet, leading to an increase in the number of
protons in the  intermediate-\pT~region.  This, however contradicts
theoretical predictions where an opposite effect was
expected~\cite{cite:Theory_Vitev}.  Alternatively, more gluon jets could
be initially produced, or {\it induced} (for example, in the radiative energy loss scenario), for the more central data.
The latter appears to be the more plausible, as the highest~\pT~data
exhibits little or no enhancement over the $pp$ data, indicating a
similar energy loss for gluons and quarks (see
Fig.~\ref{fig:BMRatio}).  Alternative approaches to explain
the phenomenon observed in the data, have also been developed.
For example, the recombination/fragmentation picture of thermal/shower
partons has had success at describing this in Au+Au data~\cite{cite:HwaRecombination}.
Further information on the relative energy loss of quark and gluon
jets can be extracted from the data by comparing the nuclear
modification factors of proton and pion data (Figs.~\ref{fig:PionRAA}~and~\ref{fig:ProtonRAA}).  
At high-\pT~(above 5~GeV/$c$), however, the two suppression factors are found to be
the same within the systematic uncertainties, suggesting a similar energy loss of quark and gluon jets
in Cu+Cu collisions.

In conclusion, new results on high-\pT~identified pion and proton spectra are presented for
several centrality bins in Cu+Cu collisions at \snn=200~GeV. The
data are found to exhibit similar systematic trends  over a wide range of transverse momenta as Au+Au collisions at the same energy with a similar number of participants. The suppression pattern observed versus the number of participants in Au+Au data is followed by the Cu+Cu data to a large degree.
The participant coverage in these Cu+Cu collisions is in a region
where the suppression effects are turning on. A detailed study of
the proton to pion ratio reveals similar systematic dependencies
to that found in Au+Au data. Specifically, the increase in proton
yield at intermediate transverse momenta persists for the much
smaller Cu+Cu system.

Further studies have shown similar suppression of protons and pions at high-\pT.
Within the  context of the connection  between the detected pions and
protons  and quark and gluon jets suggested in the introduction, these
results indicate similar partonic energy loss for both gluons and quarks.
The amount of energy loss suffered by the partons is found to
be 
\npart~dependent.  Within the experimental uncertainties,  the
suppression for different collision species is found to be invariant for the same number of participants.

\begin{acknowledgments}
We thank the RHIC Operations Group and RCF at BNL, the NERSC Center at LBNL and the Open Science Grid consortium for providing resources and support. This work was supported in part by the Offices of NP and HEP within the U.S. DOE Office of Science, the U.S. NSF, the Sloan Foundation, the DFG cluster of excellence `Origin and Structure of the Universe', CNRS/IN2P3, STFC and EPSRC of the United Kingdom, FAPESP CNPq of Brazil, Ministry of Ed. and Sci. of the Russian Federation, NNSFC, CAS, MoST, and MoE of China, GA and MSMT of the Czech Republic, FOM and NOW of the Netherlands, DAE, DST, and CSIR of India, Polish Ministry of Sci. and Higher Ed., Korea Research Foundation, Ministry of Sci., Ed. and Sports of the Rep. Of Croatia, Russian Ministry of Sci. and Tech, and RosAtom of Russia.

\end{acknowledgments}

\newpage
\appendix
\section{Monte Carlo Glauber model results for the centrality bins used in the paper}

\begin{table}[hb]
\caption{ Number of participants $N_{\rm part}$ and number of binary collisions $N_{\rm bin}$
from the Monte Carlo Glauber model calculations for different centrality bins of minimum bias  Cu+Cu collisions at  200~GeV.}
\begin{tabular}{ccc}
\\
Centrality bin&$N_{\rm part}$&$N_{\rm bin}$\\
\\
\hline
\\
0-10\%    &  $99.0^{+1.5}_{-1.2}$  &    $188.8^{+15.4}_{-13.4}$\\
\\
10-20\%   &  $74.6^{+1.3}_{-1.0}$  &    $123.6^{+9.4}_{-8.3}$\\
\\
20-40\%   &  $45.9^{+0.8}_{-0.6}$  &    $62.9^{+4.2}_{-3.7}$\\
\\
40-60\%   &  $21.5^{+0.5}_{-0.3}$  &    $22.7^{+1.2}_{-1.1}$ \\
\\
\end{tabular}
\end{table}

\begin{table}[hb]
\caption{ Number of participants $N_{\rm part}$ and number of binary collisions $N_{\rm bin}$
from the Monte Carlo Glauber model calculations for different centrality bins of minimum bias Au+Au collisions at  200~GeV.}
\begin{tabular}{ccc}
\\
Centrality bin&$N_{\rm part}$&$N_{\rm bin}$\\
\\
\hline
\\
10-20\%  & $234.6^{+8.3}_{-9.3} $     & $591.3^{+51.9}_{-59.9}$\\
\\
20-40\%  & $ 141.4^{+9.9}_{-9.5}$    & $ 294.2^{+40.6}_{-39.9}$\\
\\
40-60\%  & $ 62.4^{+8.3}_{-10.4}  $    & $ 93.6^{+17.5}_{-23.4}$\\
\\
60-80\%   & $20.9^{+5.1}_{-6.5} $     & $ 21.2^{+6.6}_{-7.9} $\\
\\
\hline
\\
40-80\%   & $41.5^{+6.9}_{-6.6}$ & $57.1^{+13.7}_{-13.3} $\\
\\
\end{tabular}
\end{table}

\begin{table}
\caption{ Number of participants $N_{\rm part}$ and number of binary collisions $N_{\rm bin}$
from the Monte Carlo Glauber model calculations for 200~GeV central triggered Au+Au collisions.}
\begin{tabular}{ccc}
\\
Centrality bin & $N_{\rm part}$ & $N_{\rm bin}$ \\
\\
\hline
\\
0-12\% & $315.7^{+5.6}_{-4.5}$ & $900.3^{+71.4}_{-63.7}$  \\
\end{tabular}
\end{table}

\end{document}